\documentclass[iop,revtex4]{emulateapj}
\usepackage{graphicx}
\usepackage{amssymb}
\usepackage{amsmath}
\usepackage{threeparttable}
\usepackage{lscape}

\mathchardef\mhyphen="2D
\newcommand{\aliii}{Al\,{\sc iii}}

\newcommand{\ovi}{O\,{\sc vi}}

\newcommand{\feii}{Fe\,{\sc ii}}

\newcommand{\siv}{S\,{\sc iv}}
\newcommand{\siiv}{Si\,{\sc iv}}
\newcommand{\siii}{Si\,{\sc ii}}

\newcommand{\pv}{P\,{\sc v}}
\newcommand{\cii}{[C\,{\sc ii}]}

\newcommand{\civ}{C\,{\sc iv}}
\newcommand{\mgii}{Mg\,{\sc ii}}
\newcommand{\angstrom}{\text{\normalfont\AA}}

\mathchardef\mhyphen="2D

\def\hi{H\,{\sc i}}

\def\cii{C\,{\sc ii}}

\def\civ{C\,{\sc iv}}
\def\niii{N\,{\sc iii}}
\def\nv{N\,{\sc v}}

\def\oiv{O\,{\sc iv}}
\def\ovi{O\,{\sc vi}}

\def\neviii{Ne\,{\sc viii}}

\def\mgii{Mg\,{\sc ii}}
\def\siiv{Si\,{\sc iv}}
\def\Siii{Si\,{\sc ii}}
\def\siii{S\,{\sc iii}}
\def\siiii{Si\,{\sc iii}}
\def\siv{S\,{\sc iv}}

\def\alii{Al\,{\sc ii}}
\def\aliii{Al\,{\sc iii}}
\def\pv{P\,{\sc v}}
\def\feii{Fe\,{\sc ii}}
\def\neviii{Ne\,{\sc viii}}
\def\mgx{Mg\,{\sc x}}
\def\nh{\vy{n}{H}}
\def\ne{\ifmmode n_\mathrm{\scriptstyle e} \else $n_\mathrm{\scriptstyle e}$\fi}
\def\Qh{\ifmmode Q_\mathrm{\scriptstyle H} \else $Q_\mathrm{\scriptstyle H}$\fi}
\def\Uh{\ifmmode U_\mathrm{\scriptstyle H} \else $U_\mathrm{\scriptstyle H}$\fi}
\def\Nh{\ifmmode N_\mathrm{\scriptstyle H} \else $N_\mathrm{\scriptstyle H}$\fi}

\def\Zsun{\ifmmode {\rm Z}_{\odot} \else Z$_{\odot}$\fi}
\def\Msun{\ifmmode {\rm M}_{\odot} \else M$_{\odot}$\fi}
\def\kms{\ifmmode {\rm km~s}^{-1} \else km~s$^{-1}$\fi}
\def\Lya{\ifmmode {\rm Ly}\alpha \else Ly$\alpha$\fi}
\def\Lyb{\ifmmode {\rm Ly}\beta \else Ly$\beta$\fi}
\def\Lyg{\ifmmode {\rm Ly}\gamma \else Ly$\gamma$\fi}
\def\Lyd{\ifmmode {\rm Ly}\delta \else Ly$\delta$\fi}
\def\neaod{\ifmmode n_\mathrm{\scriptscriptstyle AOD} \else $n_\mathrm{\scriptscriptstyle AOD}$\fi}
\def\necrit{\ifmmode n_\mathrm{\scriptstyle cr} \else $n_\mathrm{\scriptstyle cr}$\fi}
\def\ncr{\ifmmode n_\mathrm{\scriptstyle cr} \else $n_\mathrm{\scriptstyle cr}$\fi}
\def\nepi{\ifmmode n_\mathrm{\scriptscriptstyle PI} \else $n_\mathrm{\scriptscriptstyle PI}$\fi}
\def\gtorder{\mathrel{\raise.3ex\hbox{$>$}\mkern-14mu\lower0.6ex\hbox{$\sim$}}}
\def\ltorder{\mathrel{\raise.3ex\hbox{$<$}\mkern-14mu\lower0.6ex\hbox{$\sim$}}}
\newcommand{\vy}[2]{#1_{\scriptscriptstyle #2}}

\shorttitle{ }
\shortauthors{Xu et al.}
\shortauthors{}

\begin{document}


\title{A mini-BAL outflow at 900 pc from the central source: VLT/X-shooter observations}


\author{
Xinfeng Xu\altaffilmark{1,$\dagger$},
Nahum Arav\altaffilmark{1},
Timothy Miller\altaffilmark{1},
Chris Benn\altaffilmark{2}
}

\affil{$^1$Department of Physics, Virginia Tech, Blacksburg, VA 24061, USA}
\affil{$^2$Isaac Newton Group, Apartado 321, 38700 Santa Cruz de La Palma, Spain}

\altaffiltext{$\dagger$}{Email: xinfeng@vt.edu}


\begin{abstract}

We determine the physical conditions and location of the outflow material seen in the mini-BAL quasar SDSS J1111+1437 (z = 2.138). These results are based on the analysis of a high S/N, medium-resolution VLT/X-shooter spectrum. The main outflow component spans the velocity range $ -1500$ to $-3000$ km s$^{-1}$ and has detected absorption troughs from both high-ionization species: \civ, \nv, \ovi, \siiv, \pv, and \siv; and low-ionization species: \hi, \cii, \mgii, \alii, \aliii, \Siii, and \siiii. Measurements of these troughs allow us to derive an accurate photoionization solution for this absorption component: a hydrogen column density, $log(\Nh)=21.47^{+0.21}_{-0.27}$ $\rm{cm}^{-2}$ and ionization parameter, $log(\Uh)=-1.23^{+0.20}_{-0.25}$. Troughs produced from the ground and excited states of \siv \ combined with the derived \Uh \ value allow us to determine an electron number density of $log(\ne) = 3.62^{+0.09}_{-0.11}$ $\rm{cm}^{-3}$; and to obtain the distance of the ionized gas from the central source: $R=880^{+210}_{-260}$ $\rm{pc}$.

\end{abstract}






\section{INTRODUCTION}
Quasar outflows are often detected as blueshifted absorption troughs in the rest-frame of the active galactic nucleus (AGN) spectrum \citep{Hewett03, Dai08, Knigge08, Ganguly08}. 
These outflows are often invoked as agents for active galactic nuclei (AGN) feedback  \citep[e.g.,][]{Ostriker10, Ciotti10, McCarthy10, Hopkins10, Soker11, FaucherGiguere12,
Choi14, Hopkins16, Angles17, Ciotti17}. A crucial parameter needed to assess the contribution of the outflow to AGN feedback is the distance ($R$) of the outflow from the central source, which can be inferred from
excited-state absorption combined with photoionization modeling \citep[e.g.,][]{Korista08}. Using this method over the last decade, our group and other collaborations measured $R$
for about 20 AGN outflows \citep{Hamann01,deKool01,deKool02a,deKool02b,Gabel05,Moe09,Bautista10,Dunn10a,Aoki11,Arav12,Borguet12a,Borguet12b,Borguet13,Edmonds11,Arav13,Lucy14,Finn14,Chamberlain15a,Chamberlain15b}. 
These investigations located the outflows at an $R$ range of several parsecs to many kilo-parsecs. For luminous quasars, the majority of these findings were at $R$ of hundreds to thousands of
parsecs scales (see \citealt{Arav13} for details). Most of the $R$ determinations referenced above arise from singly ionized species (mainly \feii\ and \Siii). However, the majority of outflows show
absorption troughs only from more highly ionized species. Hence, the applicability of $R$ derived from singly ionized species to the majority of outflows is somewhat
model-dependent (see discussion in \S~1 of \citealt{Dunn12}). To address this issue empirically, $R$ determinations using doubly and triply ionized species are needed.

Using ground-based telescopes, the main high-ionization species with a measurable trough arising from an excited state is \siv. This ion has resonance and excited level transitions at
1062.66\angstrom/1072.97\angstrom, respectively. To measure absorption troughs from these transitions, we conducted a survey using the  VLT/X-shooter spectrograph between 2012 and 2014.
From this survey we published four $R$ determinations using \siv\ troughs \citep{Borguet13,Chamberlain15b}. Here, we present another such determination for the object SDSS J1111+1437.

The layout of this paper is as follows. In Section 2, we present the VLT/X-shooter observation of SDSS J1111+1437. In Section 3, we analyze the spectrum and extract column densities from the absorption troughs.
In Section 4, we describe the photoionization analysis that determines the ionization parameter and total hydrogen column density of the outflow. In Section 5, we analyze the density-sensitive troughs from \siv\
and \siv* to determine the electron number density \ne\ of the outflow. In Section 6, we present our distance and energetics results. Finally, we summarize our method and findings in Section 7.
\begin{figure*}[]
\label{Spec}
\centering
      \includegraphics[angle=0,clip=true,trim={0 0 10cm 1.5cm},width=1.0\linewidth]{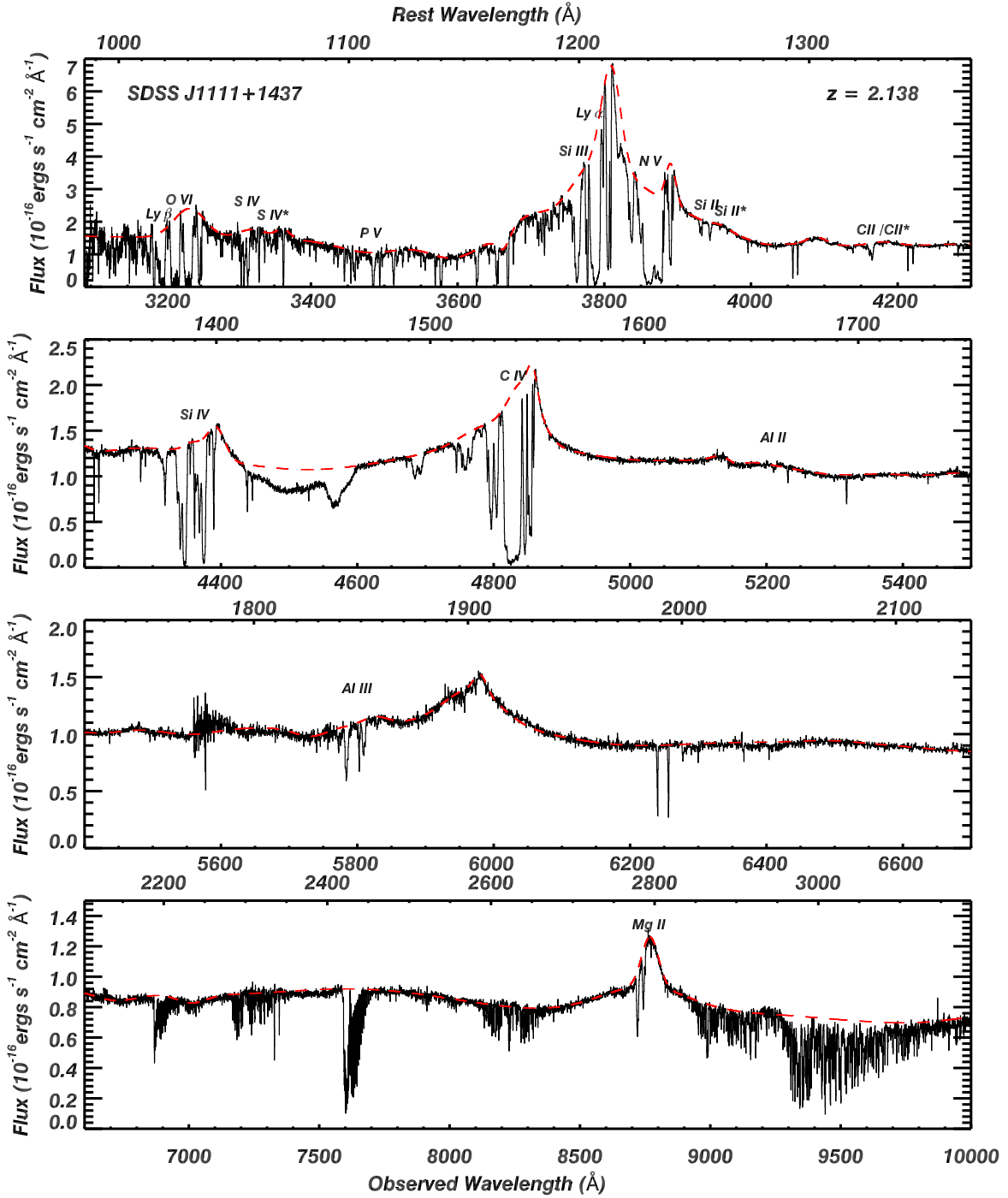}\\

\caption{VLT/X-shooter spectrum of the quasar SDSS J1111+1437 (z = 2.138). We label the ionic absorption troughs associated with the outflow and represent the unabsorbed emission model with a red dashed line (see Section 2). Narrow absorption from intervening systems appear throughout the spectrum, and terrestrial absorption from molecular $O_2$ in our atmosphere is seen near 6850$\angstrom$ and 7600$\angstrom$ (observed frame), but none of these features affect the analysis presented in this paper. We note the high-velocity \civ \  mini-BALs around 1530\angstrom, 1515\angstrom, 1495\angstrom \, as well as the even higher velocity \civ \  BAL around 1420\angstrom\ -- 1460\angstrom. However, due to the lack of diagnostic troughs, we do not analyze these four outflows in this paper. The troughs from these high-velocity systems do not affect the analysis results of the outflow component we concentrate on in this paper.} 
\end{figure*}

\section{Observation and Data Reduction}
SDSS J1111+1437 (J2000: R.A. = 11:11:10.15 , decl. = +14:37:57.1, z = 2.138) was observed with the VLT/X-shooter in January 2014 as part of our program 092.B-0267 (PI: Benn). 
X-shooter is the second-generation, medium spectral resolution (R $\sim$ 6000 -- 9000) spectrograph installed on the VLT. 
It covers a wide spectral band (3000 -- 24,000\angstrom) in a single exposure by its unique design, where the incoming light is split into three independent arms (UVB, VIS and NIR). 
The wide spectral coverage allows the detection of absorption troughs from  the ionic species \hi, \cii, \civ, \nv, \ovi, \mgii, \alii, \aliii, \Siii, \siiii, \siiv, \pv\ and \siv /\siv * (see figure \ref{Spec}). The width (1500 $\rm{km}$ $\rm{s}^{-1}$) of the \civ \  trough satisfies the definition of a mini-BAL outflow (Hamann \& Sabra 2004).

We reduced the SDSS J1111+1437 spectra using the ESO Reflex workflow \citep{Ballester11}. The one-dimensional spectra extracted were then coadded after manually performing cosmic-ray rejection on each spectrum. We present the reduced UVB+VIS+NIR spectrum of SDSS J1111+1437 in figure \ref{Spec}. 

We do not show the observed spectra between 1 $\sim$ 2.5$\micron$. As in this spectral range, we  found no absorption troughs associated with the outflow system we analyze here.   We note that, in particular, no H$\alpha$ and H$\beta$ absorption troughs were found. Existence of such troughs would suggest a higher number density for the outflow than we derive in section 5 \citep[e.g., see][]{Sun17}. Therefore, their absence is consistent with the results of our analysis.

\begin{figure*}[htp]

\centering

    	\includegraphics[angle=0,trim={0 0 0.9cm 0},clip,width=0.420\linewidth]{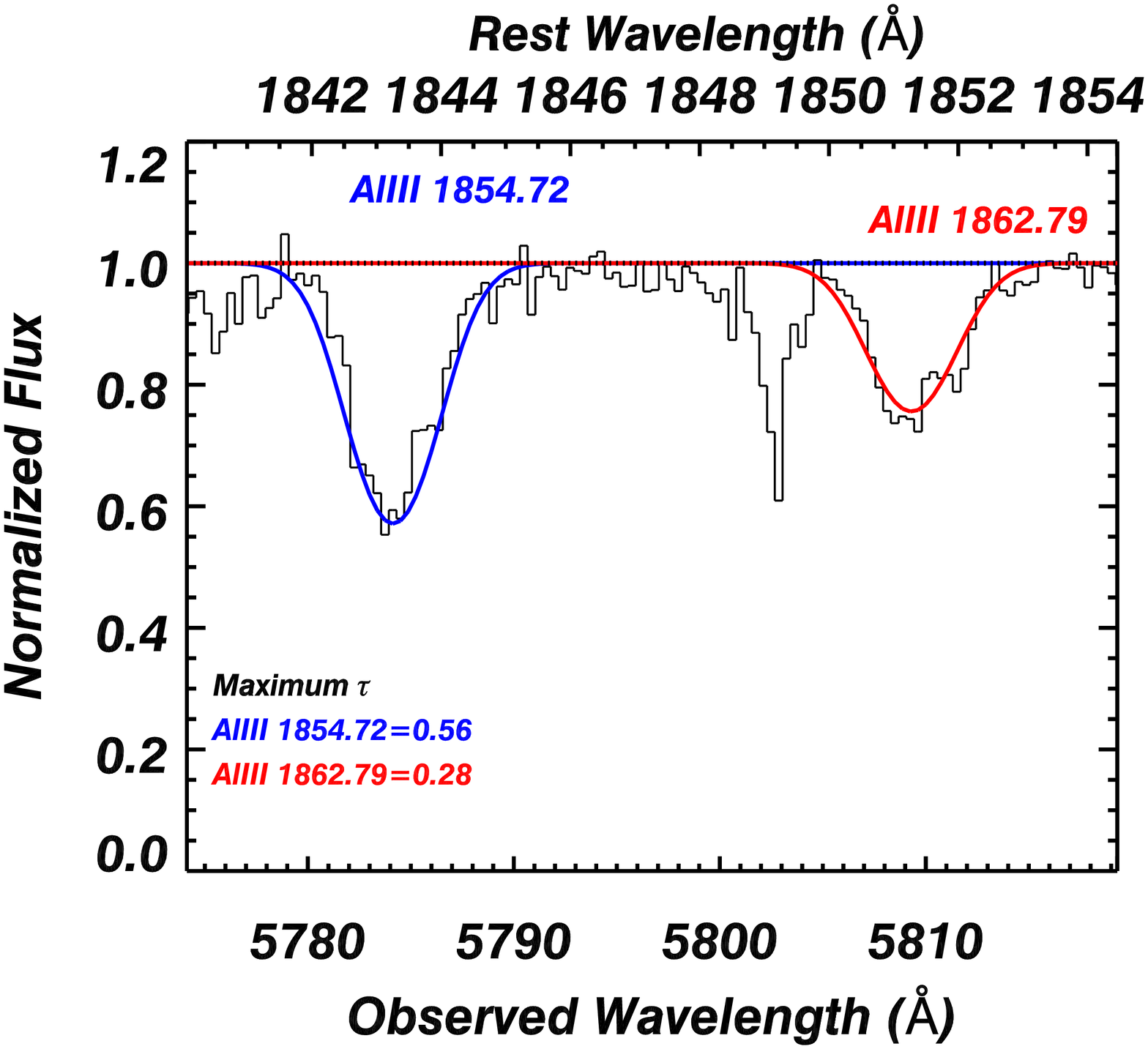} 
   	\includegraphics[angle=0,trim={0 0 0.9cm 0},clip,width=0.42\linewidth,keepaspectratio]{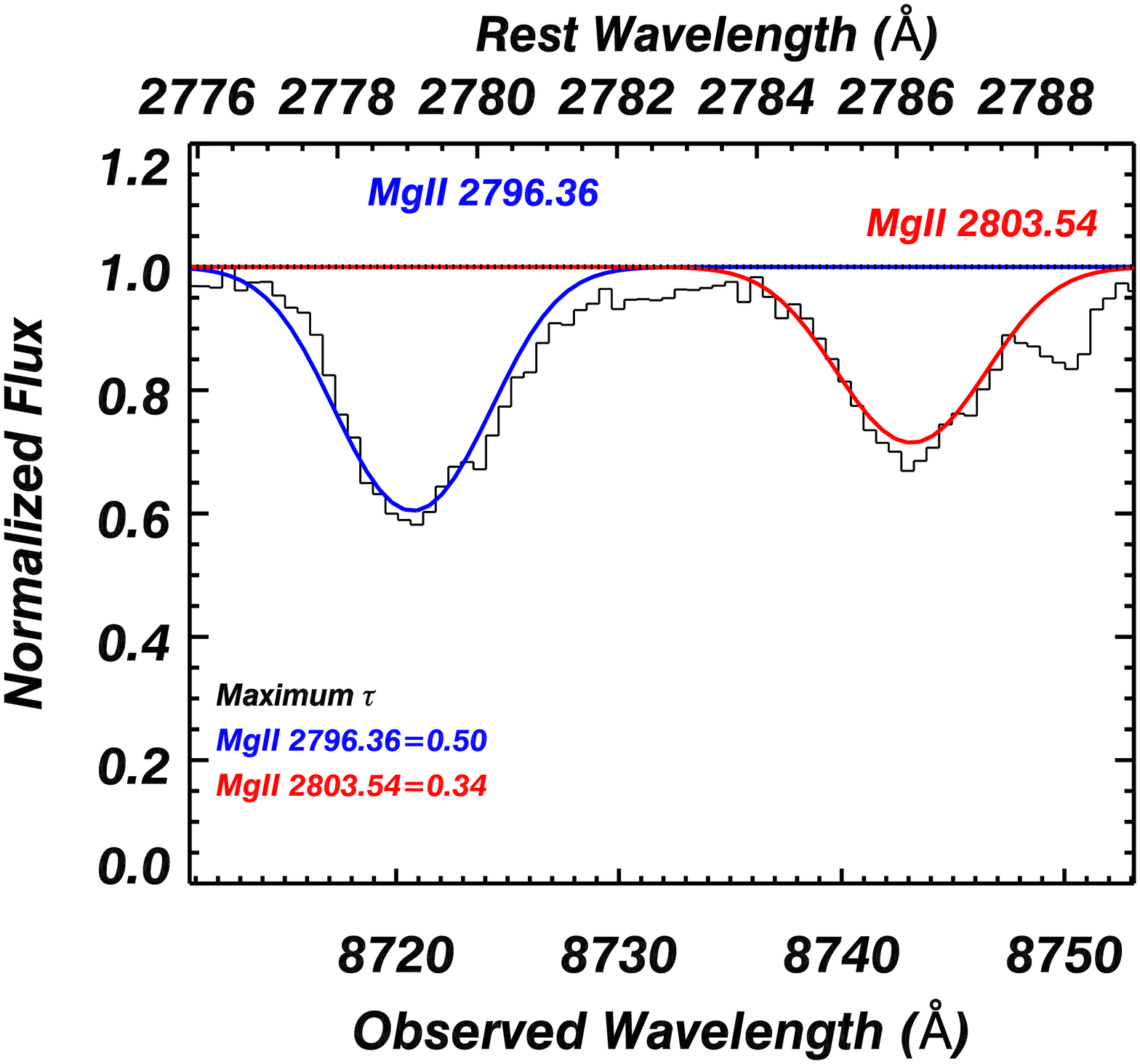}\\
    
	\includegraphics[angle=0,trim={0 0 0.9cm 0},clip,width=0.42\linewidth,keepaspectratio]{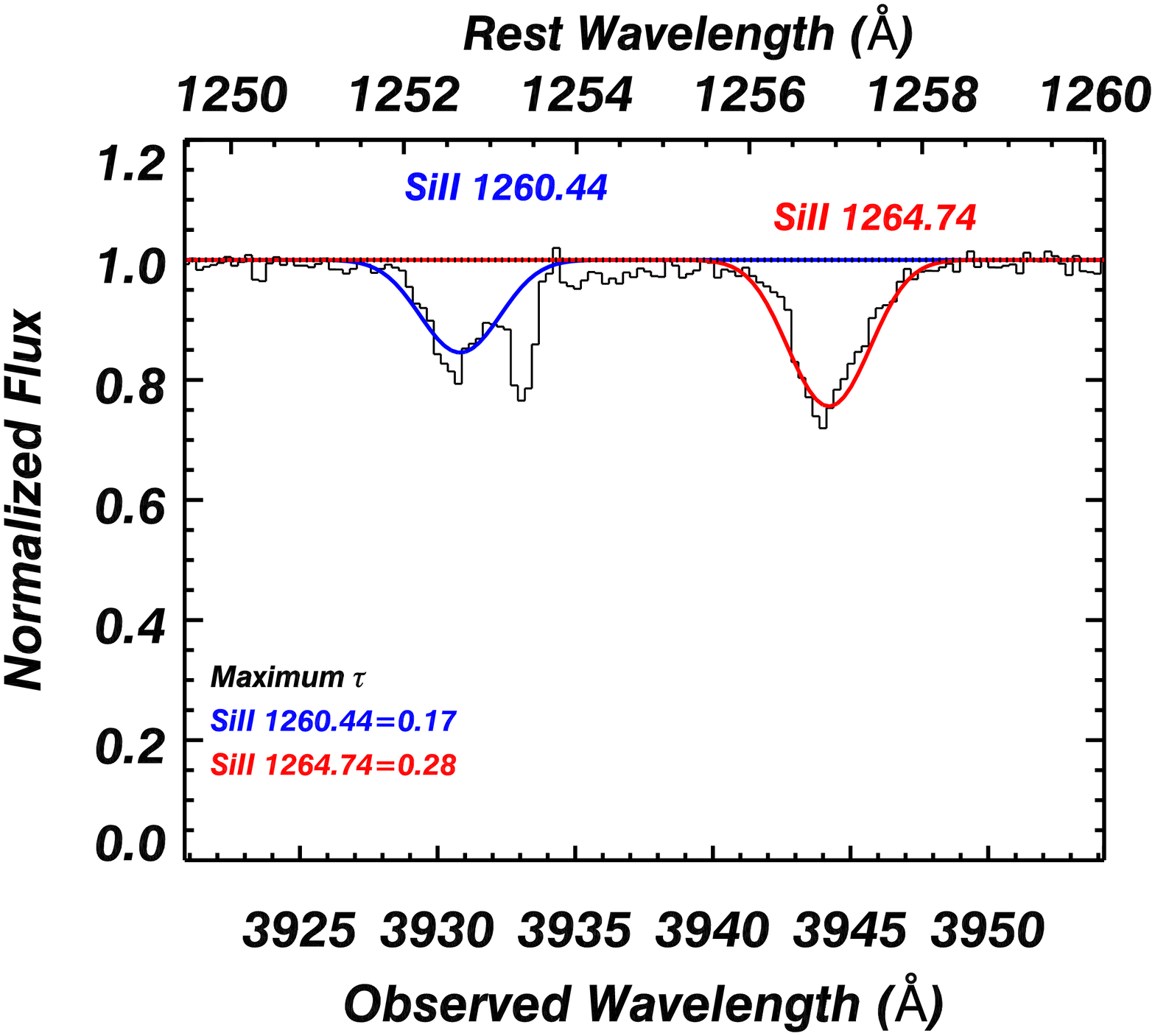}
   	\includegraphics[angle=0,trim={0 0 0.9cm 0},clip,width=0.42\linewidth,keepaspectratio]{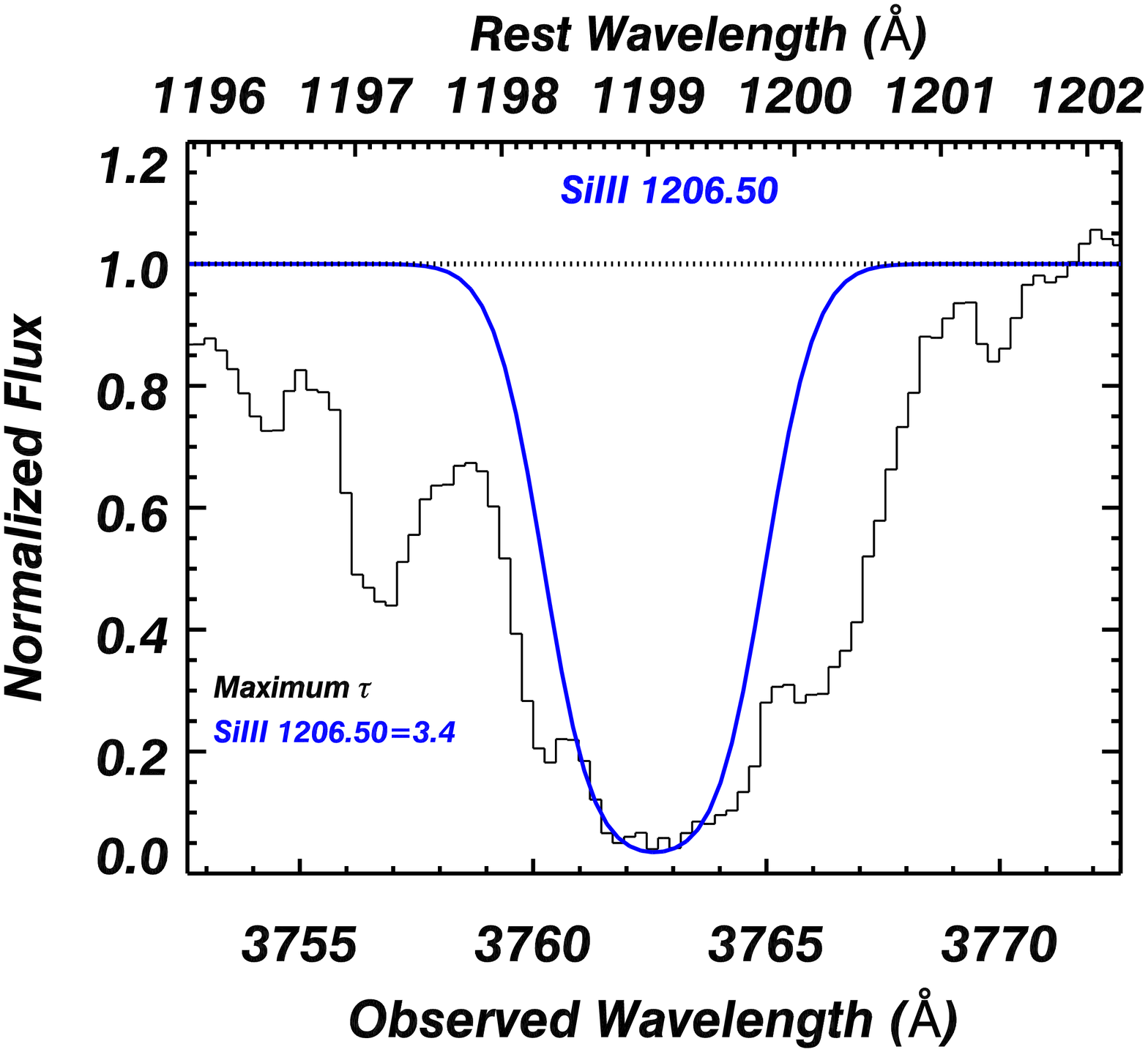}\\

    	\includegraphics[angle=0,trim={0 0 0.9cm 0},clip,width=0.42\linewidth,keepaspectratio]{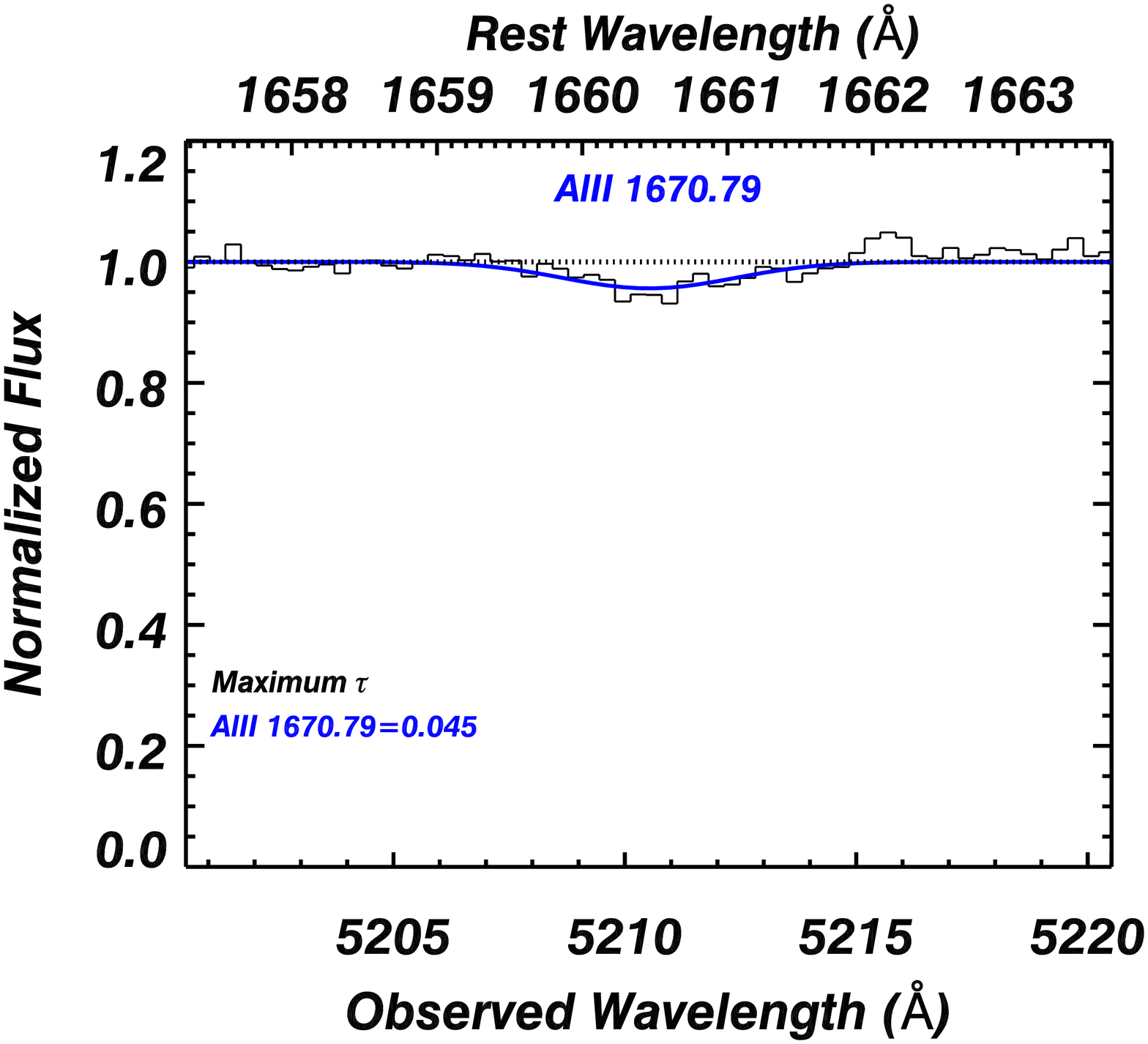}
   	\includegraphics[angle=0,trim={0 0 0.9cm 0},clip,width=0.42\linewidth,keepaspectratio]{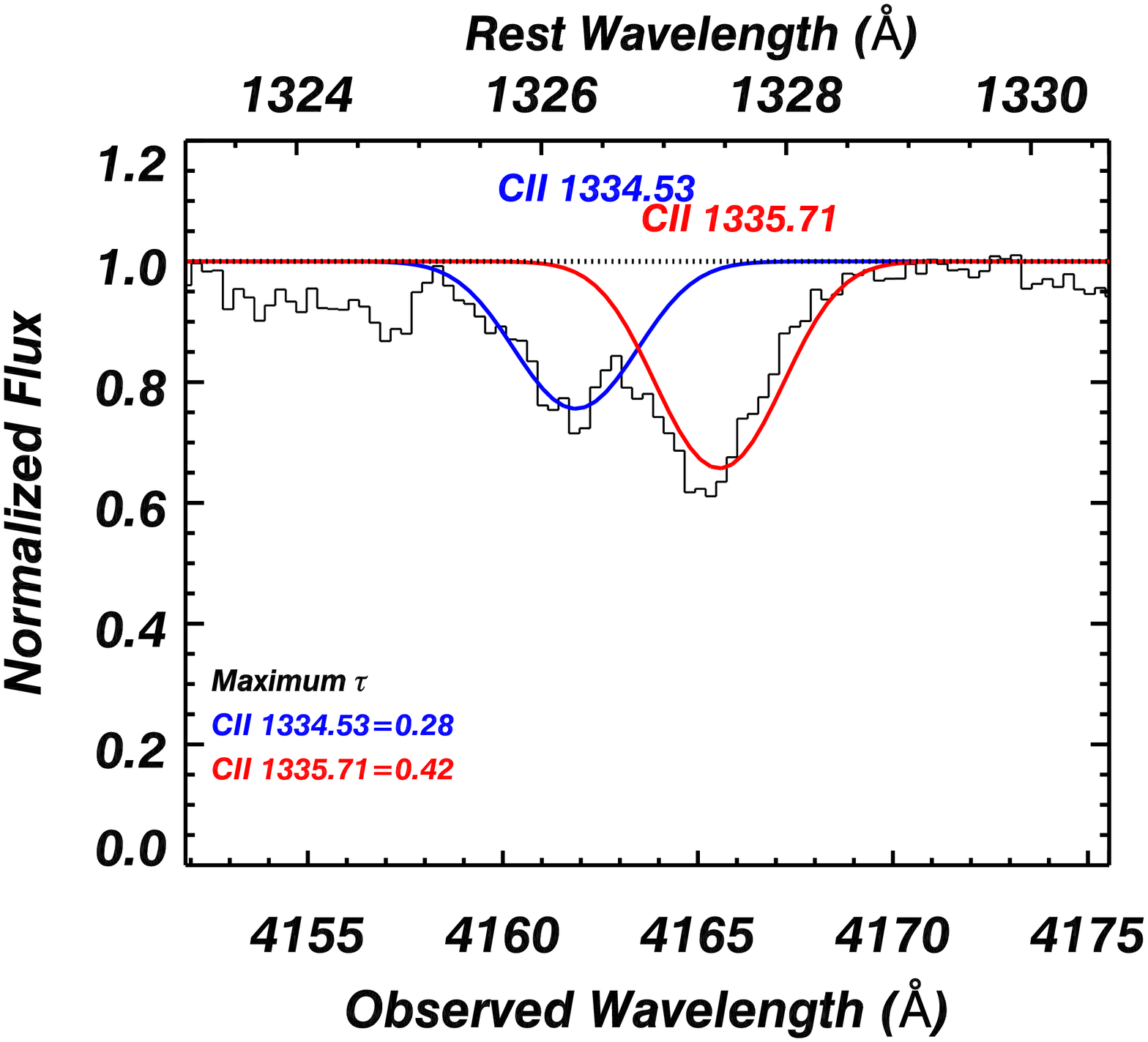}\\

\caption{Fits to the low-ionization species' absorption troughs observed in the X-shooter spectrum of SDSS J1111+1437.  The \aliii\ blue (1854.72\angstrom) trough is used for the Gaussian template fit. The Gaussian templates for the shorter- and longer-wavelength transitions are shown in blue and red, respectively. Maximum $\tau$ shown in the panels is the maximum optical depth of the fitted template.}
\label{LowIon}
\end{figure*}

\begin{figure*}[htp]
\centering
	\includegraphics[angle=0,trim={2cm 0 0.cm 0},clip,width=0.40\linewidth,keepaspectratio]{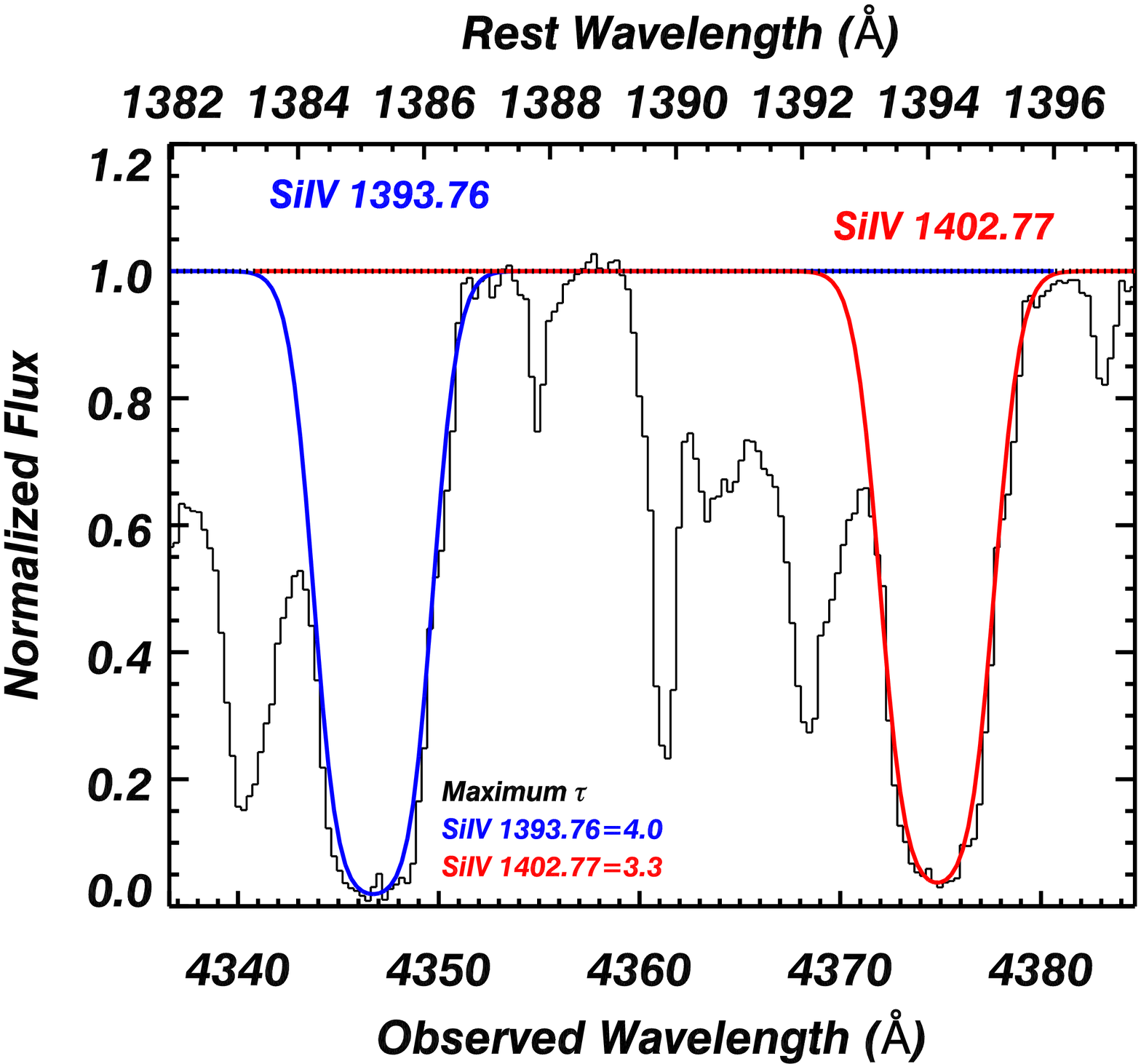}
    	\includegraphics[angle=0,trim={2cm 0 0.cm 0},clip,width=0.40\linewidth,keepaspectratio]{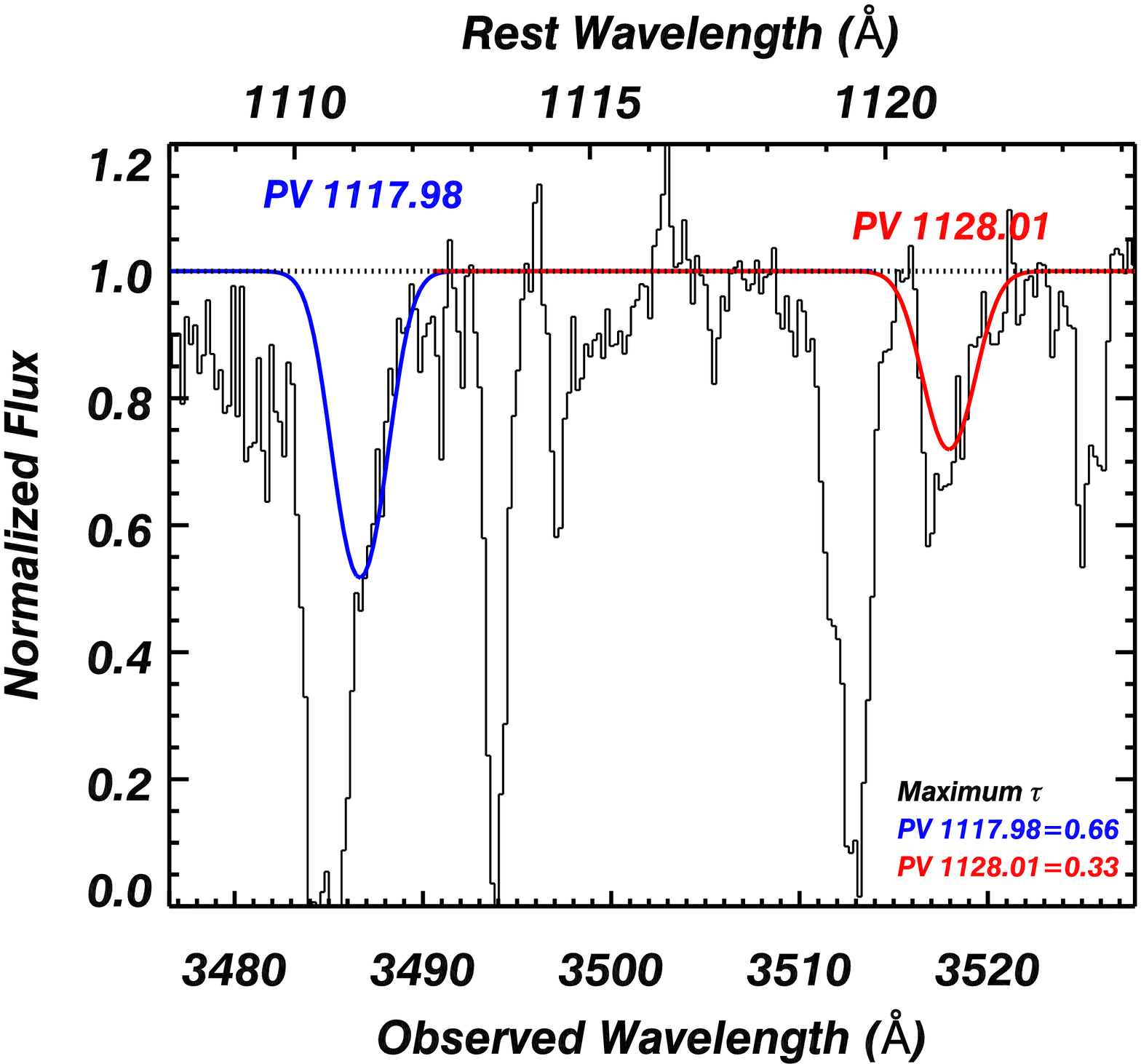}\\

	\includegraphics[angle=0,trim={2cm 0 0.cm 0},clip,width=0.40\linewidth,keepaspectratio]{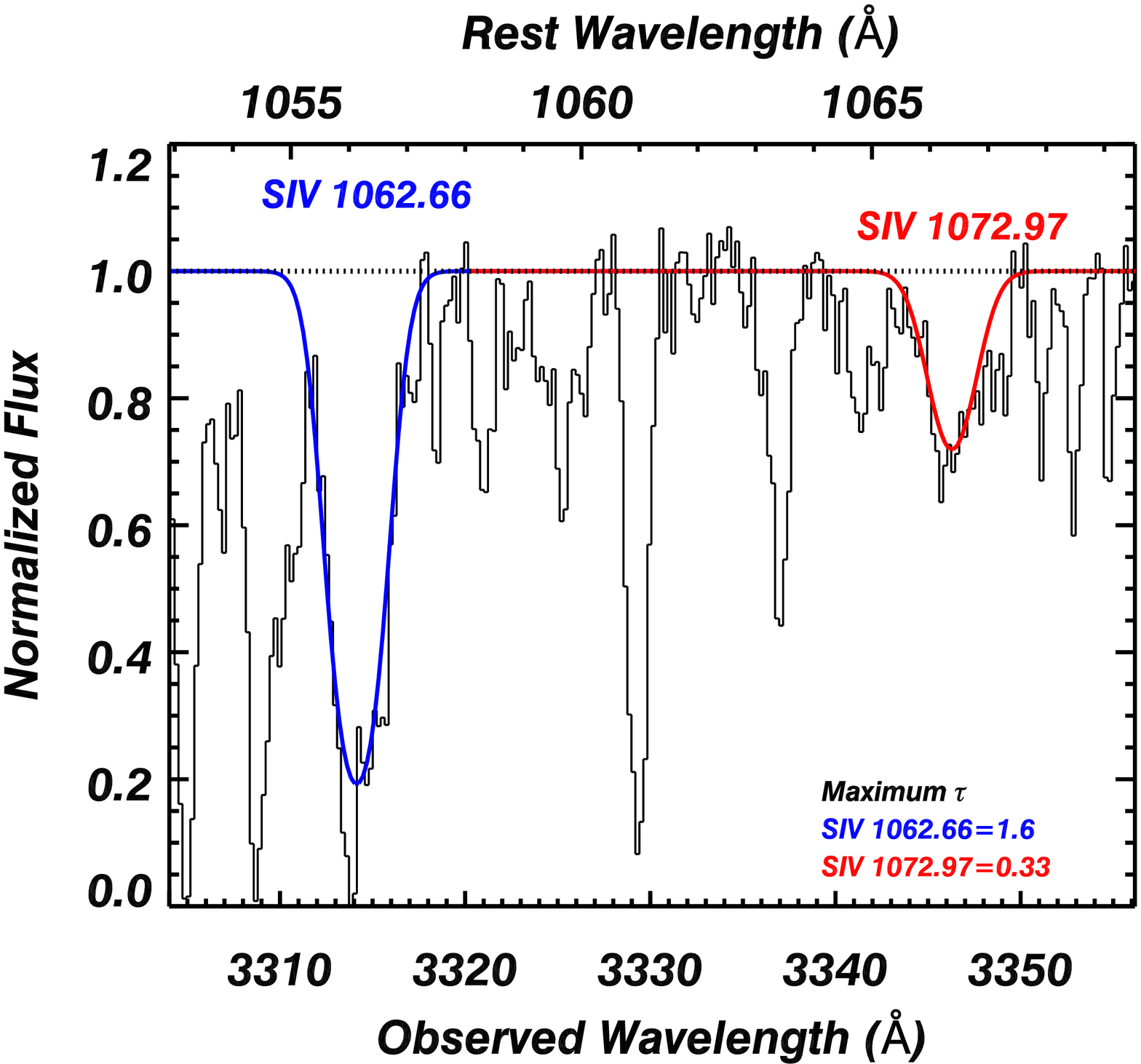}
	\includegraphics[angle=0,trim={2cm 0 0.cm 0},clip,width=0.40\linewidth,keepaspectratio]{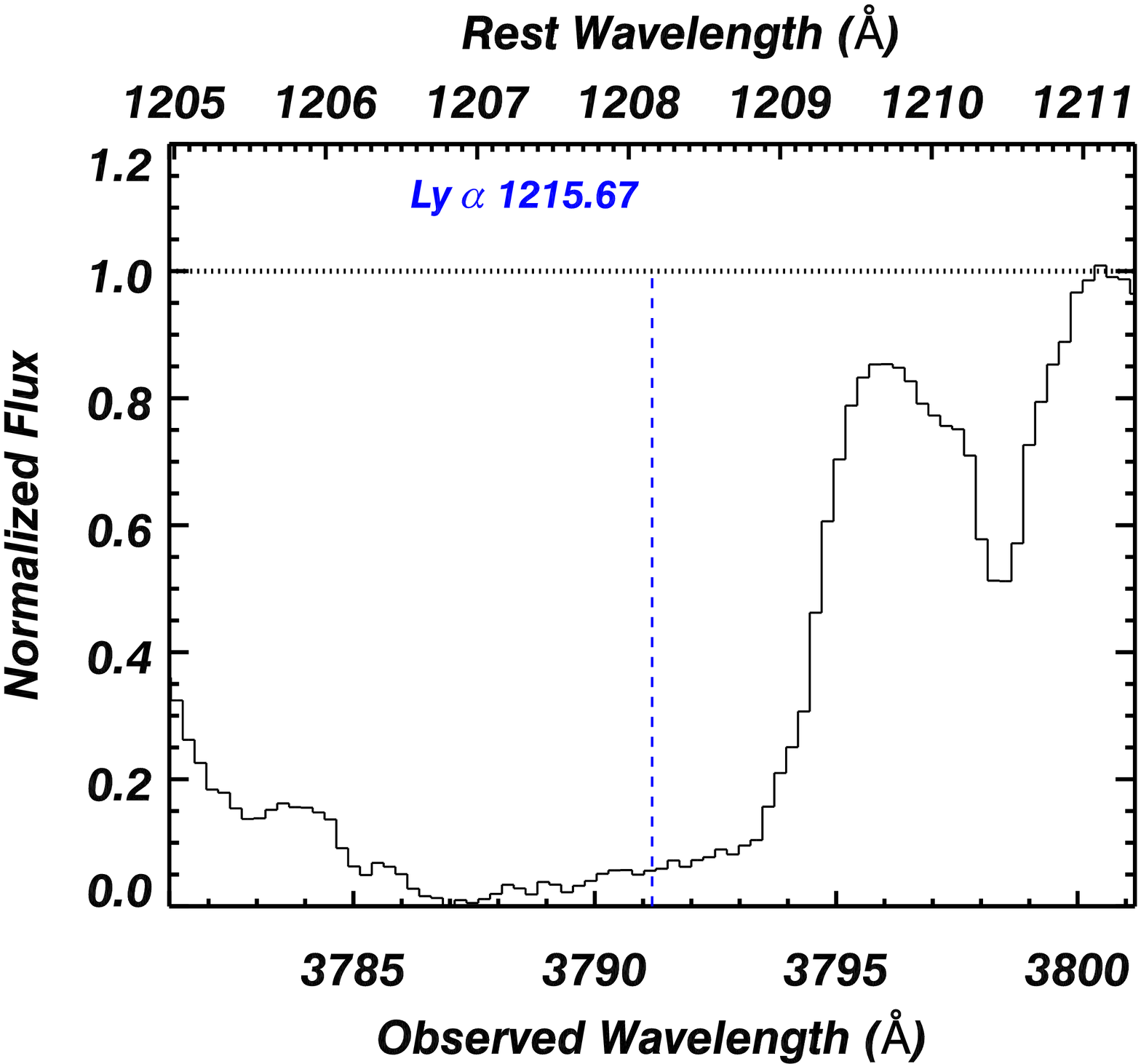}\\
	
	\includegraphics[angle=0,trim={2cm 0 0.cm 0},clip,width=0.40\linewidth,keepaspectratio]{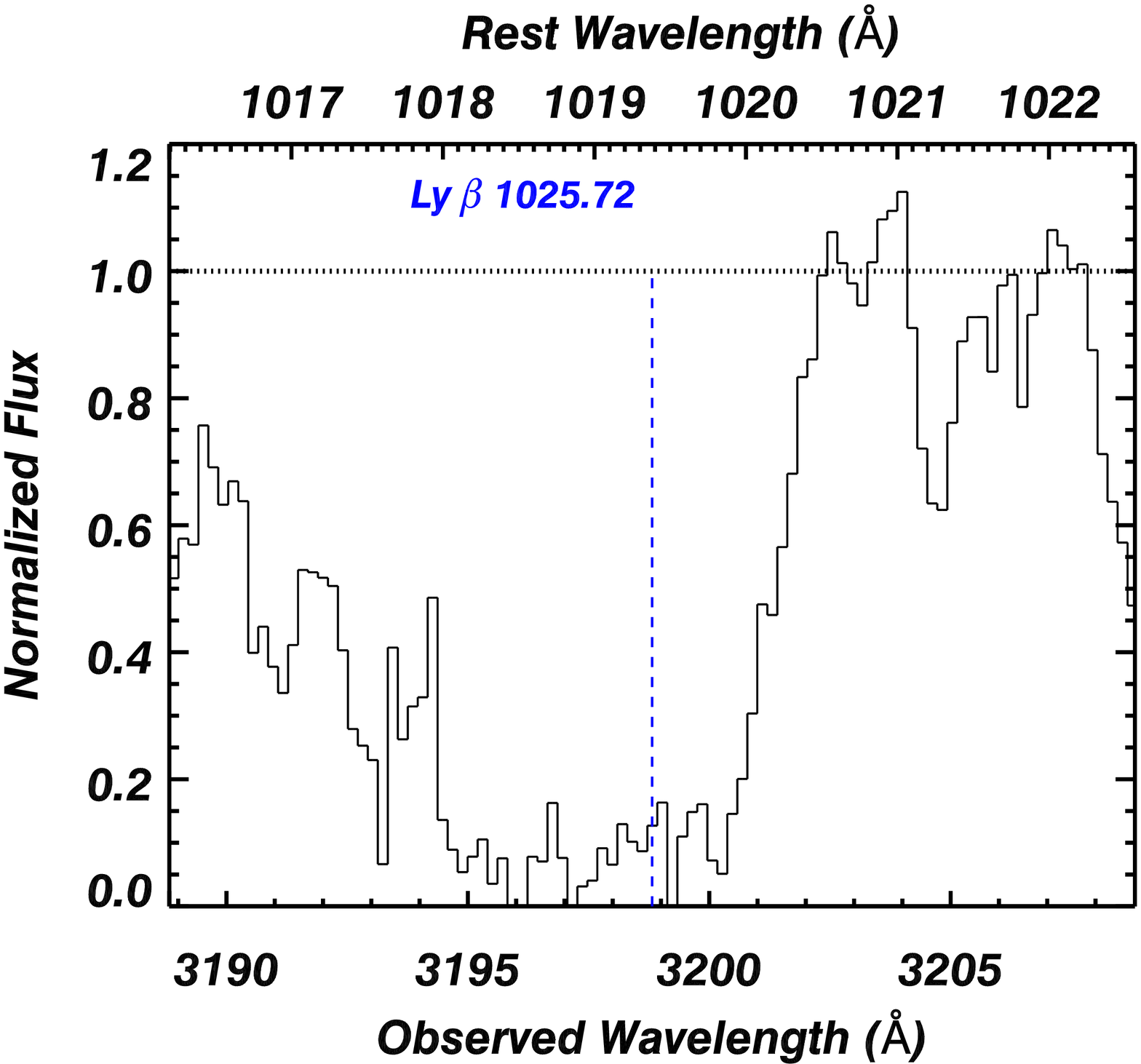}
   	\includegraphics[angle=0,trim={2cm 0 0.cm 0},clip,width=0.40\linewidth,keepaspectratio]{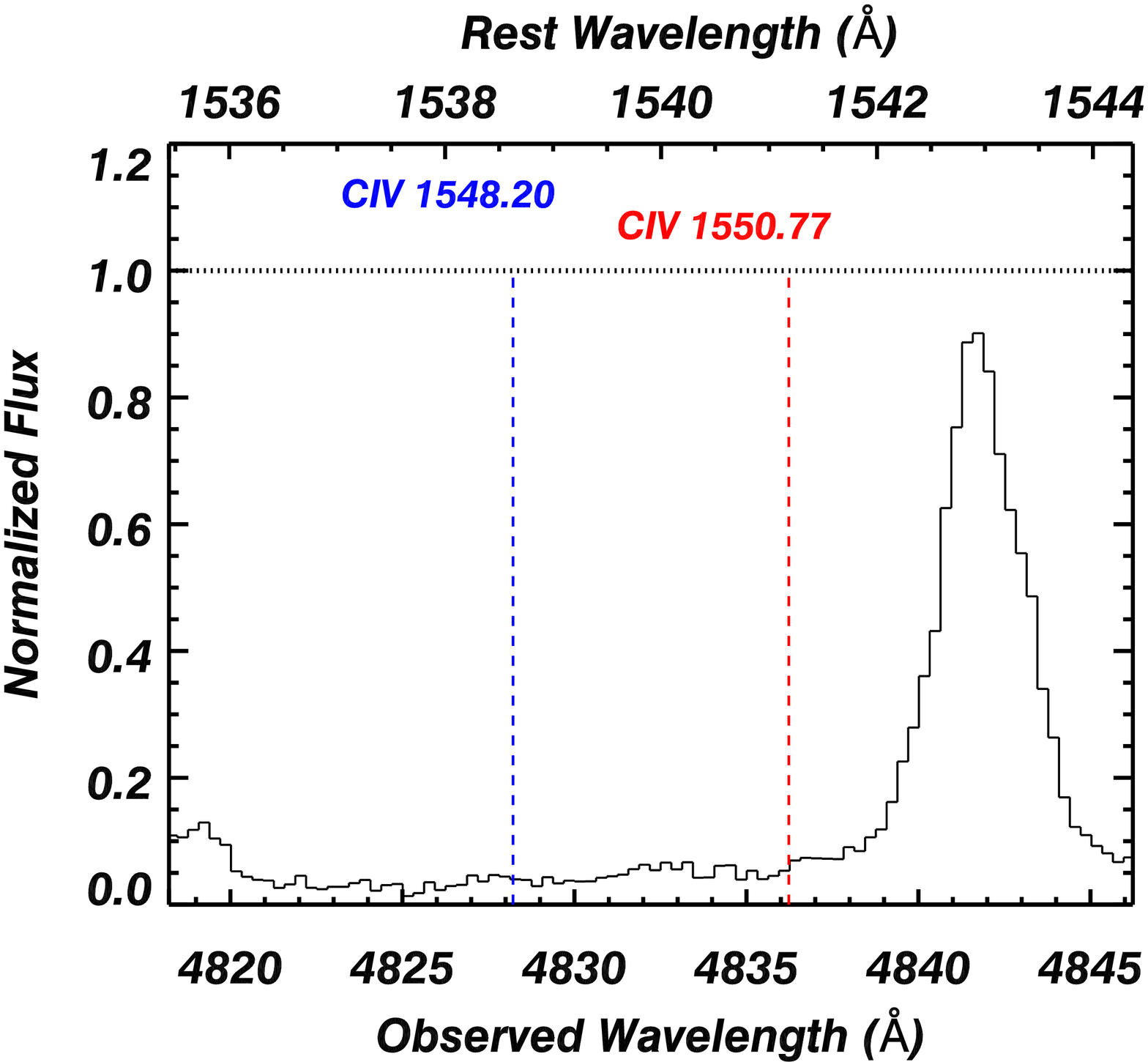}\\

	\includegraphics[angle=0,trim={2cm 0 0.cm 0},clip,,width=0.40\linewidth,keepaspectratio]{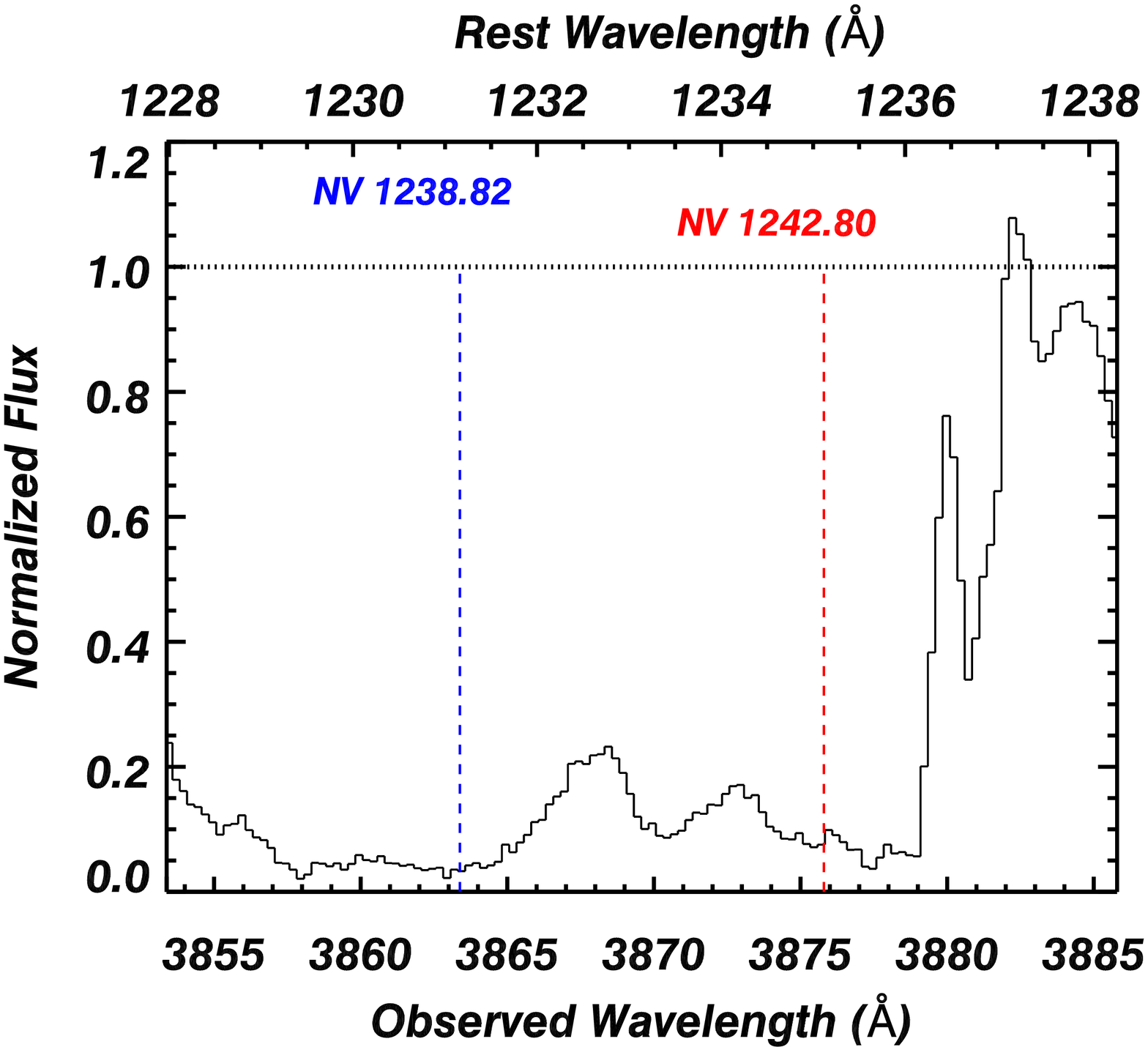}
	\includegraphics[angle=0,trim={2cm 0 0.cm 0},clip,width=0.40\linewidth,keepaspectratio]{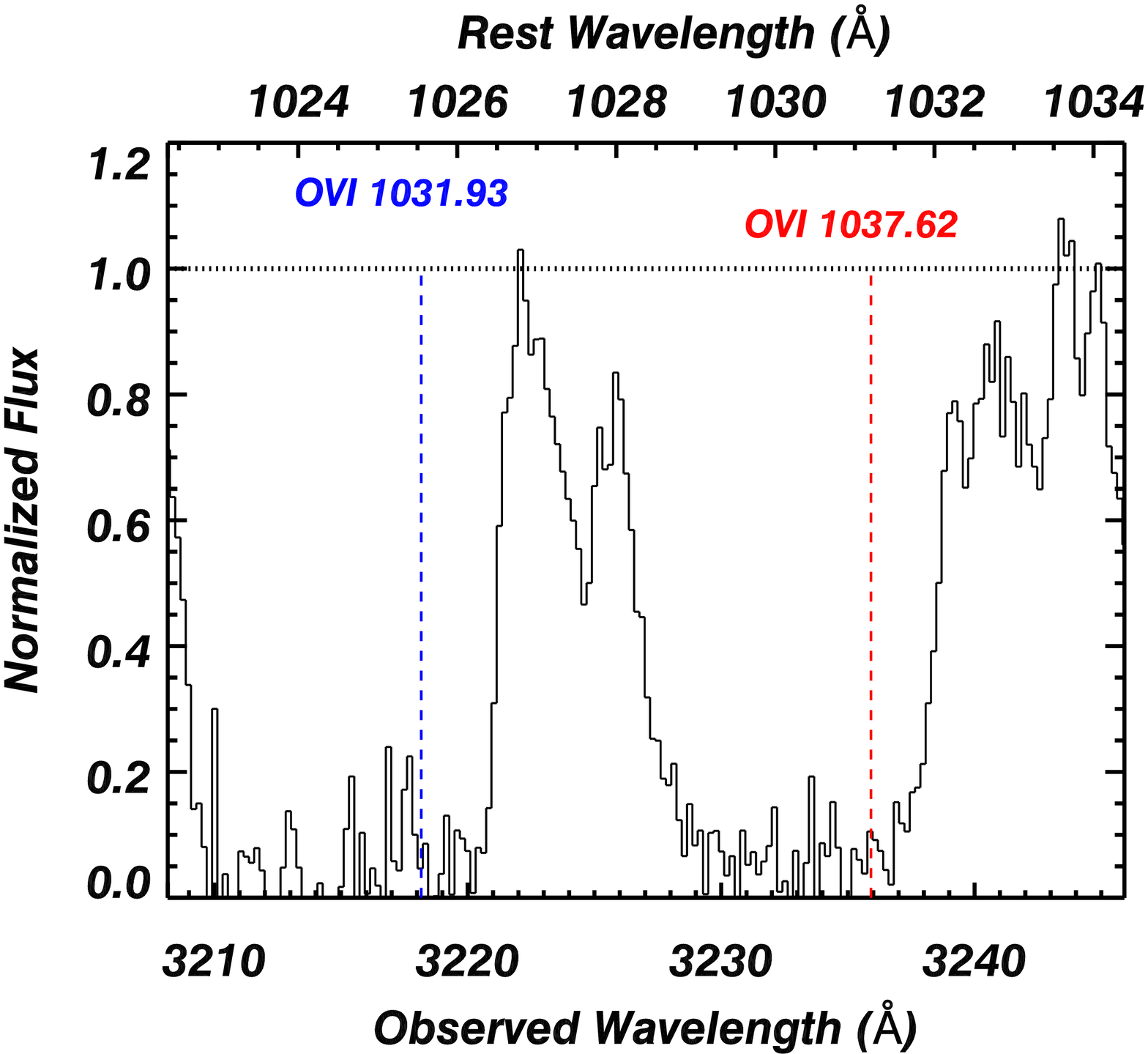}\\

\caption{Same as figure \ref{LowIon} but for the high-ionization species and the saturated \Lya\ and \Lyb\ troughs. The \siiv \ red (1402.77\angstrom) trough is used for the Gaussian template fit. For the saturated line troughs where template fitting is not adopted, the column density integration range is --2200 $<$ v$<$ --1500 km s$^{-1}$.}
\label{HighIon}
\end{figure*}

\begin{deluxetable}{c l l l l}[t]
\tablewidth{0.5\textwidth}
\tabletypesize{\small}
\setlength{\tabcolsep}{0.02in}
\tablecaption{Measured Column Densities\label{tab:colTable}}
\tablehead{
\colhead{Ion} & \colhead{AOD$^{a}$} & \colhead{PC} & \colhead{PL} & \colhead{Adopted}
\\
\colhead{(1)} & \colhead{(2)} & \colhead{(3)} & \colhead{(4)} & \colhead{(5)} 
}

\startdata
\hi   		&  $>$9800				&  			&  					&  $>9800_{-1960}$\\
\cii  		&  $>$320 				&  			&  					&  $>320_{-64} $\\
\civ   		&  $>$3100				&   			& 					&  $>3100_{-620}$ \\
\nv   		&  $>$5100 				&  			& 	 				&  $>5100_{-1020}$ \\
\ovi   		&  $>$6600 				&  	 		&	 				&  $>6600_{-1320}$ \\
\mgii		&  $>44$				&  $53.6^{+0.4}_{-0.4}$	& $58^{+2.8}_{-1.7}$ 			&  $53.6^{+7.2}_{-10.1}$	 \\
\alii   	&  $>2.8$				&  			& 				 	&  $>2.8_{-0.56}$ \\
\aliii 		&  $>48$				&  $56.6^{+1.6}_{-2}$	& $68^{+8.6}_{-5.2}$ 			&  $56.6^{+20}_{-8.6}$\\
\Siii 		&  $>30$				&  			&					&  $>30_{-8.5}$ \\
\siiii 		&  $>200$				&  			&					&  $>200_{-40}$ \\
\siiv 		&  $>700$				&  			&					&  $>700_{-140}$ \\
\pv  		&  $>144$ 				&  $145^{+16}_{-16}$ 	& $155^{+18}_{-20}$			&  $145^{+28}_{-16}$ \\
\siv  		&  $>3400^{+450}_{-450}$  		&  			& 					&  $>3400^{+450}_{-450}$ \\
\siv * 		&  $>800^{+125}_{-125}$	 		&   			& 		 			&  $>800^{+125}_{-125}$ \\
\enddata

\tablecomments{\\
$^{a}$ The velocity integration range for the AOD method is from --2200 to --1500 km s$^{-1}$.
Columns (2),(3),(4): The integrated column densities for the three absorber models, in units of $10^{12} \rm{cm}^{-2}$. The errors are computed from photon statistics (S/N). 
Column (5): Adopted values for the photoionization analysis. The computation of the adopted error bars is described in Section 3.2.
}

\label{table_ions}
\end{deluxetable}

\section{Spectral Fitting}

\subsection{Unabsorbed Emission Model}
	Generally, the UV unabsorbed emission source in AGNs can be modeled by two components: a continuum that is often described by a power law and emission lines that are usually modeled by a set of ad hoc Gaussian profiles. These emission lines can be divided into broad emission lines (BELs) and narrow emission lines. \\
	
	We fit the continuum with a power law and where the fit is poor (1090\angstrom \  -- 1170\angstrom \  and 2100\angstrom \  -- 2300\angstrom \ rest-frame), we correct it by a cubic spline. The power law is in the form of $F(\lambda) = F_{1100}(\lambda/1100)^\alpha$, where $F_{1100}=1.5\times10^{-16} \rm{erg s}^{-1}\rm{cm}^{-2} 		\angstrom^{-1}$ is the observed flux at 1100$\angstrom$ (rest-frame) and $\alpha$ = -0.704. Figure 1 shows our full unabsorbed emission model fitted to the data. 
	
\subsection{Modeling of Blended Troughs and Column Density Measurements}
An outflow system is identified by absorption troughs from different lines that cover a similar velocity range. In SDSS J1111+1437, the analyzed outflow spans --2200 $<$ v $<$ --1500 km s$^{-1}$. To derive the ionic column density, the simplest model is the apparent optical depth (AOD) method, where $\tau(\lambda) \equiv -ln(I(\lambda))$, $I(\lambda) = F_{obs}(\lambda)/F_{0}(\lambda)$ is the residual intensity and $F_{obs}$ is the observed flux. The optical depth, $\tau$, is then used to calculate the associated column density, $N_{ion}$
(e.g. Equation (9) in \citealt{Savage91}). However, AOD analysis usually gives a lower limit for column densities due to non-black saturation in the troughs of AGN outflows \citep[e.g.,][]{Arav08,Borguet12a, Borguet13, Chamberlain15b, Arav18}. Therefore, for singlet lines we use the AOD measurement only as a lower limit.\\ 

For doublet and multiplet line troughs, we can use the partial-covering (PC) and power-law (PL) absorption models to overcome the non-black saturation problem \citep{Arav99a,deKool02b,Borguet12a,Chamberlain15b}.
However, self-blending exists in the \cii, \civ\ and \nv\ troughs, and blending with the \Lya\ forest affects the \siv, \pv\ and \ovi\ troughs. 
For these cases, we can only use the template fitting method \citep{Moe09, Borguet13,Chamberlain15b,Arav18}.
The main assumption in the template fitting method is that ions with a similar ionization potential will have similar optical depth profiles as a function of velocity. 
In this case, unblended troughs can be used as a template to fit a Gaussian profile in optical depth. Then the blended troughs can be fitted with this Gaussian profile by scaling the Gaussian's depth but leaving the shape and centroid velocity unchanged.\\

We chose the unblended \aliii \ 1854.72$\angstrom$  \ trough as the template for the low-ionization troughs, including \cii, \mgii\ and \Siii, while high-ionization troughs, like \civ, \nv, \ovi\ and \siv,  were fitted by the \siiv \ 1402.77$\angstrom$ template. 
The main reasons for choosing these two troughs are: 1. the wide velocity separation between their blue and red component of the doublet along with the width of the outflow prevents self-blending; 2. these two troughs are not contaminated by other intervening absorption lines; and 
3. the low-ionization species have similar ionization potentials to \aliii, while the high-ionization species have ionization potentials closer to that of \siiv. For both of these templates, we keep the centroid velocity and width fixed and change the scaling of the template to get the best fitting for other troughs. Then we extract ionic column densities using the methods described in \cite{Chamberlain15a}, using the AOD, PC, and PL methods on the fitted templates. See figures \ref{LowIon} and \ref{HighIon} for more details on the template and table \ref{table_ions} for each ion's column density. 

The final column density value adopted for each ion follows these rules. If a PC measurement exists, we choose it as the adopted value. The upper error bar is set to be the PL's upper bound minus the PC column density value, while the lower error is the PC measurement minus the AOD column density value. In this way, we take into account the systematic error that is due to different absorption models \citep{Chamberlain15b}. If the trough is from a singlet, we choose the AOD measurement as a lower limit. The lower error of the the adopted AOD value is fixed at 20\%. This includes the systematic error of the emission model in the template fitting. 

Examining figures 2 and 3 demonstrates the advantages of the template fitting approach. First, the template fitting to the \aliii, \mgii, \siii\ and \cii\ troughs in Figure 2 show good matches to the data, which increase our confidence that these troughs are close to the apparent optical depth case, and therefore do not suffer from much non-black saturation. The minor substructure within these absorption features that are not well matched by smooth Gaussian templates, contributes less than 5\% error to the column densities of the above troughs.
Second, as mentioned above, the template fits allow us to extract column densities for troughs in the Ly$\alpha$ forest, by ignoring the contaminating troughs. This attribute is crucial for the diagnostic \pv\ and \siv\ troughs.


\subsection{The Low-Ionization Troughs}
The \aliii \ doublet troughs are well separated without any contamination, and the optical depth ratio between the \aliii \ blue and red troughs is 2:1. This means that the troughs are not saturated. We fit a single Gaussian to the \aliii \ blue trough at 1854.72$\angstrom$, and we used that Gaussian profile as the template for all other low-ionization troughs.

The \Lya \ and \Lyb \ line troughs are highly saturated; the AOD column densities are 3200 and 9800 ( in units of $10^{12}\ \rm{cm}^{-2}$), respectively. The different values indicate saturation, and we therefore use the higher value (\Lyb's) as a lower limit for the hydrogen column density.

The \mgii \ troughs are well separated but the red component's right wing is contaminated by the \mgii \ blue trough from another outflow system (see the feature at 8750\angstrom\ [observed frame] in figure \ref{LowIon} top right panel). By adopting the \aliii \ template fitting, we can factor out the contamination of this outflow system. The ratio between the blue and red components of \mgii \ is not exactly 2:1, which means they are mildly saturated. Therefore, for this situation we use the PC value. 

\cii, \alii, \Siii\, and \siiii\ are all singlets, thus the PC and PL models are not applicable and the saturation is undetermined. Therefore, we use the AOD method to get lower limits on their column densities.

\subsection{The high-ionization troughs}
In the \siiv \ doublet, we see several different outflow systems spanning  --3000 $<$ v $<$ --1500\ km s$^{-1}$. The deepest absorption trough at --2200 $<$ v $<$ --1500 km s$^{-1}$ is the one coinciding with the low-ionization and the \siv \ troughs. Although the multiple components blend with each other to some extent, one Gaussian profile can fit the blue and red components of \siiv \ between  --2200 $<$ v $<$ --1500 km s$^{-1}$ quite well. The lower velocity wings are uncontaminated and well fitted by the Gaussian template. In \cite{Arav18}, it has been shown that the \siiv \ absorption trough itself can work as a good template for \siv \ troughs (see their figure 4). Here, we use the Gaussian profile fitted on the \siiv\ trough (1402.77\angstrom) as the template for \pv \  and \siv /\siv*, as they are blended with the \Lya \ forest (see figure \ref{HighIon} for more details).

The troughs from the \pv \ doublet are important diagnostics in AGN outflows since the low abundance of phosphorous ($\sim 10^{-3}$ that of carbon in solar abundance; \citealt{Lodders09}) makes the \pv\ trough less likely to be saturated. Here, the blue wing of \pv \ (1117.98\angstrom) \ is clearly contaminated by \Lya\ forest absorption lines. However, the template fits the red wing of this trough quite well, thus allowing us to measure its full column density. In figure \ref{HighIon}, the template fitting shows that the red and blue components of \pv \ have a 1:2 optical depth ratio. Therefore, they are not saturated and the AOD measurement yields reliable column densities.

The high-ionization troughs \civ, \nv\ and \ovi, are highly saturated. We adopt their AOD results as the lower limits for their column densities.

\section{Photoionization Analysis}
Photoionization is the dominant ionization mechanism in AGN outflows \citep{Borguet12a, Borguet13, Chamberlain15a}. We solve the ionization and thermal balance equations with the spectral synthesis code Cloudy (version c17.00), which is designed to simulate conditions in interstellar matter under a broad range of conditions \citep{Ferland17}.
\subsection{Photoionization Solution}
We ran a grid of photoionization simulations using the spectral energy distribution, UV-soft SED \citep{Dunn10a} and assumed solar metallicity. We  varied the ionization parameter log(\Uh) between -5.0 and 3.0 in steps of 0.05 dex, with the stopping criterion that the ratio of the proton density to total hydrogen density (log(\Nh)) equals 0.1. This stopping criterion ensures that the gas zones generated by Cloudy cover a large range of hydrogen column densities for each \Uh, while optimizing the computation time when a region is near the hydrogen ionization front. Then we compared the predicted column densities from the grid model to our measurements for each ion. Each colored contour in figure \ref{Photon} represents all possible solutions for the measured column density of a given ion. We then performed a $\chi^2-$minimization of the difference between the model and measured column densities to find the best-fit solution: $log(\Nh)=21.47^{+0.21}_{-0.27}$ $\rm{cm}^{-2}$ and $log(\Uh)=-1.23^{+0.20}_{-0.25}$. This solution is indicated by the red `X' surrounded by the 1$\sigma$ confidence level contour (see figure \ref{Photon} for more details). 

\subsection{The Possibility of Multi-ionization Phases}

Our photoionization analysis implicitly assumes that the outflowing material has one ionization phase.  However, there are known cases of quasar outflows where at least two ionization components are needed \citep{Arav01,Gabel05,Arav13}.
Therefore, we need to address the ramification of such a possibility on our current analysis.

\begin{figure}[hb]
\centering
\includegraphics[width=1.0\columnwidth,keepaspectratio,angle=0,trim={0 0 0 0},clip=true]{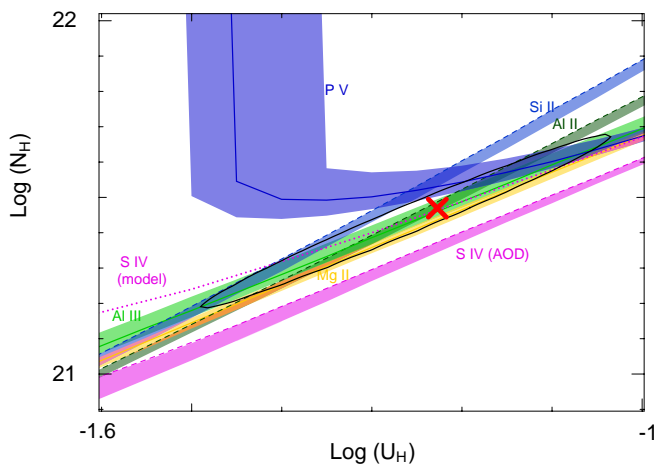}
\caption{Grid-models showing the photoionization solution. For the studied outflow component in SDSS J1111+1437, we use the UV-soft SED \citep{Dunn10a} and assume solar metallicity for the gas. Each colored contour represents the region of models ($U_H$, $N_H$) that predicts a column density consistent with the observed column density for that ion. Solid lines represent measurements, while dashed lines represent lower limits. The dotted line labeled ``\siv\ (model)" is where we match the \siv \ column density to the photoionization solution (see section 5 for more details). The photoionization solution is the red `X' and the $1\sigma$ error contour is the black ellipse. All other ion measurements in table \ref{table_ions} that are not shown here are lower limits below the plot range, but are consistent with the photoionization solution.}
\label{Photon}
\end{figure}

We start by noting that the photoionization solution we present in figure \ref{Photon}, fits the measured column density from all the troughs in our data.   These troughs arise from low-ionization species such as \alii\  to the high-ionization \pv. The good fit of a single ionization solution to all these measurements argues that we do not miss material with a lower ionization potential than \pv. 

However, there is the good possibility that a higher ionization phase exists and would have been revealed if we could observe troughs from higher ionization species \citep[e.g., \neviii\  and \mgx][]{Arav13}.  Such a phase can contribute a larger \Nh\ than the lower ionization phase \citep[see][]{Arav13},
increasing the derived mass flux and kinetic luminosity linearly with the total \Nh\  (see equation  3). 
In contrast, the determination for the distance of the outflow from the central source (R) will probably not change appreciably (see section 6).

\subsection{Dependency on SED and Metallicity}

\begin{figure}[hb]

\centering
\includegraphics[width=1.0\columnwidth,keepaspectratio,angle=0,trim={0 0 0 0},clip=true]{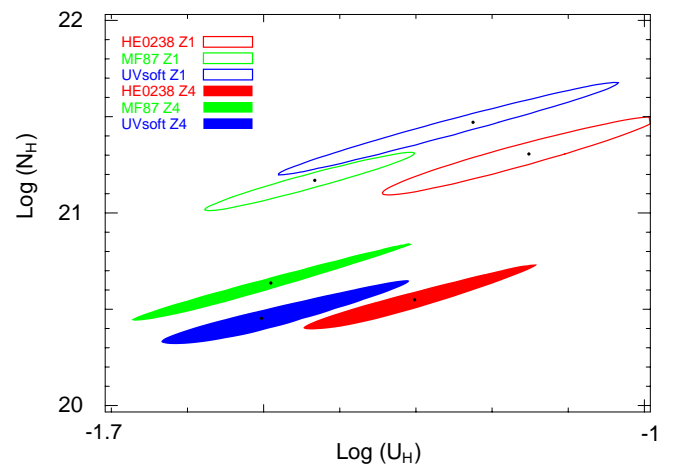}%
\caption{Grid-models showing the photoionization solution for three SEDs (HE0238, MF87 and UVsoft, see the descriptions in \cite{Arav13}) and two metallicities: solar metallicity (Z1) and four times solar metallicity (Z4), for a total of six models.}
\label{zSED}
\end{figure}

In order to show the influence the choice of SED and metallicity value have on the photoionization results, we consider different pairs of SEDs and metallicities following the approach of \cite{Arav13} and \cite{Chamberlain15b}. There are three different SEDs used in \cite{Arav13} along with two different metallicities for a total of six different cases. In figure \ref{zSED}, we present the results of each case, where the $\chi^2-$contours are shown in different colors. GASS10 \citep{Grevesee10} solar metallicity is used for Z1, and GASS10 with the scaling of \cite{Hamann93} is used  for four times solar metallicity (Z4). An increase in metallicity lowers the \Nh\ value by roughly the same amount. This happens because the metals' column densities are fixed, and when the metallicity is increased, the column density ratios between the metals and hydrogen will increase. Therefore, the total hydrogen column density drops. The different SEDs and metallicities spread the solution over 0.3 and 1 dex for \Uh \ and \Nh, respectively.  

\begin{deluxetable*}{ l c c c c c c c c c c c}[htb!]
\tablewidth{\textwidth}
\tabletypesize{\small}
\setlength{\tabcolsep}{0.02in}
\tablecaption{Comparison with Other Energetic Quasar Outflows\label{tab:compareTable}}
\tablehead{
 \colhead{Object}& \colhead{log($L_{Bol}$)} &\colhead{BH Mass} & \colhead{v} & \colhead{log(\Uh)} & \colhead{log(\Nh)} & \colhead{log(\ne)} & \colhead{R} & \colhead{$\dot{M}$} & \colhead{Log $\dot{E_{k}}$} & \colhead{ $\dot{E_{k}}/L^{d}_{Edd}$}
\\
\\ [-2mm]
 \colhead{}& \colhead{erg s$^{-1}$} & \colhead{log(M/$M_{\bigodot}$)}  & \colhead{km s$^{-1}$} & \colhead{log(cm$^{-2}$)}& \colhead{log(cm$^{-2}$)} & \colhead{log(cm$^{-3}$)}& \colhead{pc} & \colhead{$M_{\odot}$ yr$^{-1}$} & \colhead{log(erg s$^{-1}$)} & \colhead{$\%$} 
}

\startdata
	HE 0238-1904$^{a}$  		&47.2 	& -	&-5000	&-1.8$^{+1}_{-0.4}$	&20.7$^{+0.09}_{-0.1}$	&4.5$^{+0.2}_{-0.2}$	&1700$^{+1200}_{-1200}$	&69$^{+50}_{-50}$	&45.4$^{+0.3}_{-0.6}$ 	&1.6$^{+1.3^{(3)}}_{-1.2}$\\
	J0318-0600$^{b}$		&47.6	&9.6	&-4200	&-3.1			&19.9			&3.3			&6000			&120		 	&44.8			&0.13	 \\
	J0831+0354$^{c}$		&46.9  	&8.8	&-10,800	&-0.3$^{+0.5}_{-0.5}$	&22.5$^{+0.5}_{-0.4}$	&4.4$^{+0.3}_{-0.2}	$&110$^{+30}_{-25}$	&410$^{+530}_{-220}$	&46.2$^{+0.4}_{-0.3}$	&14$^{+18}_{-7.7}$\\
	J0838+2955$^{b}$		&47.5	&9.0	&-4900	&-1.9$^{+0.2}_{-0.2}$	&20.8$^{+0.3}_{-0.3}$	&3.8$^{+0.2}_{-0.2}$	&3300$^{+1500}_{-1000}$	&300$^{+2100}_{-120}$	&45.4$^{+0.2}_{-0.2}$	&2$^{+1.2}_{-0.8}$ \\
	J1106+1939$^{c}$		&47.2	&8.9	&-8250  &-0.5$^{+0.3}_{-0.2}$	&22.1$^{+0.3}_{-0.1}$	&4.1$^{+0.02}_{-0.4}$	&320$^{+200}_{-100}$	&390$^{+300}_{-10}$	&46.0$^{+0.3}_{-0.1}$	&12$^{+11}_{-0.3}$\\
	J1111+1437$^{c}$		&46.9	&9.2	&-1860 	&-1.23$^{+0.20}_{-0.25}$&21.47$^{+0.21}_{-0.27}$&3.62$^{+0.09}_{-0.11}$	&880$^{+210}_{-260}$	&55$^{+10}_{-11}$	&43.8$^{+0.07}_{-0.1}$	&0.03$^{+0.005}_{-0.006}$\\
	J1206+1052$^{d}$  		&47.6	&9.0  	&-1400	&-1.82$^{+0.12}_{-0.12}$&20.46$^{+0.2}_{-0.2}$	&3.03$^{+0.06}_{-0.06}$ &840$^{+60}_{-60}$	&9$^{+3}_{-3}$		&42.8$^{+0.15}_{-0.15}$	&0.001$^{+0.0005}_{-0.0025}$ \\

\vspace{-2.2mm}
\enddata

\tablecomments{
\\
(1). All solutions assumed the UV-soft SED with metallicity Z = $Z_{\bigodot}$, except for SDSS J0318-0600, the reddened UV-soft SED with Z = $7.2Z_{\bigodot}$ is assumed; for SDSS J0838+2955, the modified MF87 SED with Z = $Z_{\bigodot}$ is assumed; and for SDSS J1106+1939, the UV-soft SED with Z = $4Z_{\bigodot}$ is assumed.\\
(2). \ne\ is derived from:\\ 
$^{a}$ high-ionization: \oiv */\oiv.\\ 
$^{b}$ low-ionization: \Siii */\Siii. \\
$^{c}$ high-ionization: \siv */\siv. \\
$^{d}$ \siii */\siii\ and \niii */ \niii.\\
(3) For HE0238-1904, we report $\dot{E_{k}}/L_{Bol}$.\\
\indent\textbf{References:} HE 0238-1904: \cite{Arav13}; SDSS J0318-0600: \cite{Dunn10a}; SDSS J0831+0354: \cite{Chamberlain15a}; SDSS J0838+2955: \cite{Moe09}; SDSS J1106+1939: \cite{Borguet13}; SDSS J1111+1437: this  work; SDSS J1206+1052: \cite{Chamberlain15b}.
}
\label{tableCompare}
\end{deluxetable*}

\section{The density-sensitive troughs: \siv \ and \siv*}
\begin{figure}[]

\centering
\includegraphics[angle=0,clip=true,width=1\linewidth,keepaspectratio]{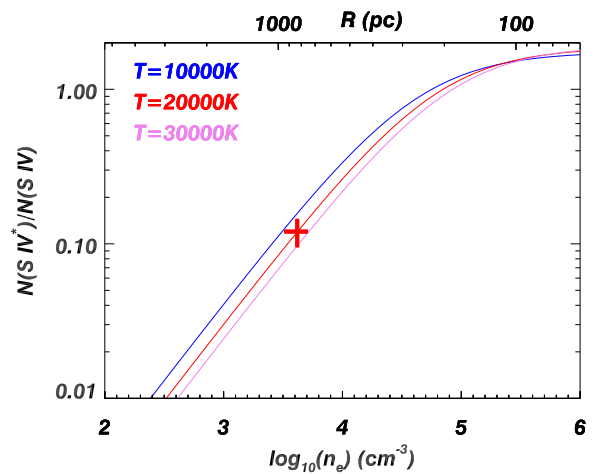}
\caption{The \siv*/\siv\ column density ratio versus \ne\ from Chianti following Equation (1). The red cross matches the solution from the photoionization plot (see figure \ref{Photon}). To show the sensitivity of the derived \ne\ to temperature, we include the curves for three different temperatures. The mean temperature of the S IV gas in our Cloudy simulation from the photoionization solution is 20,000K. The top axis is the outflow's distance from the central source using equation (\ref{Eq:ionP}) (see section 6 for more details).}
\label{SIV}
\end{figure}

The \siv* energy level is populated by electrons collisionally excited from the ground state by free electrons. Therefore, the ratio between the column densities of \siv * and \siv \ can be used as a diagnostic for the electron number density, \ne. The temperature of the \siv \ zone predicted by the Cloudy simulation from our adopted photoionization solution is 20,000K. From \cite{Arav18}, we have: 
\begin{equation}
\label{func:ne}
n_e \simeq n_{cr}\left [\frac{2N(S\  IV)}{N(S\ IV^{*})}e^{-\Delta E/kT}-1 \right ]^{-1}
\end{equation}

where $n_{cr}=6.3\times10^{4}\ \rm{cm}^{-3}$ is the critical density for the excited/ground states at  T=20,000K, N(\siv) and N(\siv *) are the column densities for the ground and excited states of \siv, respectively, $\Delta$E is the energy difference between the levels, T is the temperature in Kelvin, and k is the Boltzmann constant. In practice, we compare our measured \siv */\siv \  column density ratio to those predicted by a grid of collisional excitation models calculated from the Chianti 7.1.3 atomic database \citep{Landi13} to obtain the number density of the outflowing gas (see figure \ref{SIV}). 

\indent In order to obtain \ne\ from equation (1), we need reliable values for N(\siv) and N(\siv*). The AOD values for N(\siv) and N(\siv*) are given in table \ref{table_ions} and their combined value is plotted in figure \ref{Photon}, as the strip labeled ``S IV (AOD)". However, this ``S IV (AOD)" contour  does not intersect the photoionization solution. This is not surprising since N(\siv) and N(\siv*) are calculated using the AOD method and are therefore lower limits. To obtain the total N(\siv) that matches the solution, we ran a Cloudy model for the \Nh\ and \Uh\ values of the solution shown in figure \ref{Photon}. The predicted N(\siv) $+$ N(\siv*) value is  N(\siv$_{model-tot})=7140^{+950}_{-950}\times10^{12}\ \rm{cm}^{-2}$, which we then plot as the ``S IV (model)" contour in figure \ref{Photon}. Next, we need to obtain separate values for N(\siv) and N(\siv*). We can do so under the plausible assumption 
that the \siv\ and \siv* troughs share the same covering factor C(v), and then use the partial-covering factor formalism \citep[e.g.,][]{Dunn10b} to solve for both, which yields
N(\siv*$_{model})=860^{+130}_{-130}\times10^{12}\ \rm{cm}^{-2}$. In this situation, the deeper \siv \ trough is much more saturated (around twice the AOD value) than the shallower \siv* trough (only 10\% above its AOD value). We use these model values to recalculate the  N(\siv*)/N(\siv) ratio (see figure \ref{SIV}), and obtain $log(\ne) = 3.62^{+0.09}_{-0.11}$ $\rm{cm}^{-3}$. Finally, in the high-ionization region where \siv\ is abundant, $\ne \simeq 1.2 \nh$. Therefore, we derive a hydrogen number density of log($\nh$)=$3.54^{+0.08}_{-0.09}$  $\rm{cm}^{-3}$.

\section{Outflow distance and Energetics}

A photoionized plasma is characterized by the ionization parameter \Uh:

\begin{equation}
\label{Eq:ionP}
\Uh=\frac{\Qh}{4\pi R^2 \nh c}
\end{equation}

where $\Qh =5.7\times 10^{56}\ s^{-1}$ is the source emission rate of hydrogen ionizing photons, R is the distance of the outflow to the central source, $\nh$\ is the number density of hydrogen, and c is the speed of light.
We can obtain a measurement for R by solving equation (2), which yields the following result: the outflow of SDSS J1111+1437 is located at $R=880^{+210}_{-260}$ $\rm{pc}$ from the central source, where the errors come from adding in quadrature the errors for \Uh\ and $\nh$.

We note that an additional higher ionization phase in the outflowing material would not affect the $R$ determination appreciably.  This is because such a higher ionization phase will have only a negligible amount of \siv. Therefore, the gas containing the \siv/\siv* we measure is almost entirely associated with the ionization solution we present in figure \ref{Photon}.

Assuming the outflow is in the form of a thin partial shell, its mass flow rate ($\dot{M}$) and kinetic luminosity ($\dot{E_{k}}$) are given by \citep{Borguet12a}:
\begin{equation}\label{eq:1}
\begin{split}
\dot{M}\simeq 4\pi \Omega R\Nh \mu m_p v =55^{+10}_{-11}M_{\odot}\ \rm{year}^{-1}
\end{split}
\end{equation}

\begin{equation}\label{eq:2}
\begin{split}
\dot{E}\simeq \frac{1}{2} \dot{M}v^2= 6^{+1.1}_{-1.2}\times10^{43}\rm{erg}\ \rm{s}^{-1}
\end{split}
\end{equation}

where R is the distance of the outflow from the central source, $\Omega=0.08$ is the global covering factor for outflows showing \siv \ absorptions\citep{Borguet13}, $\mu$ = 1.4 is the mean atomic mass per proton, $m_p$ is the proton mass, \Nh\ is the absorber's total hydrogen column density, 
and $v$ is the radial velocity of the outflow. The results in equations (3) and (4) are calculated from the photoionization solution shown in figure \ref{Photon}, which uses the UV-soft SED and assumes solar metallicity \citep{Dunn10a}.
As evident from the error ellipse in figure \ref{Photon}, \Nh\ and \Uh\ are correlated. Since $R$ is a function of \Uh, \Nh\ and $R$ are also correlated, and we took this into account in calculating the error quoted in equation (3). 

\subsection{Comparison with Other Objects}

We compare our results with several other outflows which have distance and energetic determinations in the literature (see Table \ref{tableCompare}). As shown in the Introduction, $\dot{E_{k}}$ values exceeding 0.5\% (Hopkins \& Elvis 2010) or 5\% (Scannapieco \& Oh 2004) of the Eddington luminosity are viewed to be potentially significant for AGN feedback. For SDSS J1111+1437, the mass of the supermassive black hole (SMBH) is estimated to be $log(M_{BH}/M_{\odot}) \sim9.2$ [from the virial theorem and the full width half maximum (FWHM) of the \civ \ BEL using equation (3) of \cite{Park13}]. The Eddington luminosity is then $L_{Edd}\sim2\times10^{47} \rm{erg}\ \rm{s}^{-1}$. Combined with the result from equation (4), our determination of $\dot{E_{k}}/L_{edd}= 0.03\%$ shows that SDSS J1111+1437 does not have significant AGN feedback from this outflow. The other four higher-velocity outflows (see figure 1) might have higher $\dot{E}_k$. However, we lack the diagnostic troughs to measure their R and \Nh.

\section{Summary}

We presented an analysis of an outflow seen in quasar SDSS J1111+1437 based on observations from the VLT/X-shooter. Our results are summarized as follows:

1. We analyzed the outflow component that spans the velocity range --1500 to --2200 km s$^{-1}$ with a velocity centroid of -1860 km s$^{-1}$ in the quasar's rest-frame. This outflow component shows a variety of absorption troughs from both high and low-ionization species, for which we derive ionic column densities (see table \ref{table_ions}).

2. From the density-sensitive troughs of \siv\ and \siv*, we determined the electron number density as $log(\ne) = 3.62^{+0.09}_{-0.11}$ $\rm{cm}^{-3}$. In the high-ionization region where \siv \ is abundant, $\ne \simeq 1.2 \nh$, which gives a hydrogen number density of log($\nh$)=$3.54^{+0.08}_{-0.09}$  $\rm{cm}^{-3}$.

3. Using the derived column densities, we presented the photoionization plots in the log(\Nh) -- log(\Uh) phase space to find the photoionization solution. We also tested the dependency on the choice of SED and metallicity, and chose the UV-soft SED with solar metallity as the representative result, which gave a column density for hydrogen of $log(\Nh)=21.47^{+0.21}_{-0.27}$ $\rm{cm}^{-2}$ and an ionization parameter of $log(\Uh)=-1.23^{+0.20}_{-0.25}$.

4. The distance of the outflow was determined to be $880^{+210}_{-260}$ $\rm{pc}$ from the central source, along with a mass flow rate of $\dot{M} = 55^{+10}_{-11}M_{\odot}\ \rm{year}^{-1}$ and kinetic luminosity at 0.03\% of the Eddington luminosity.

\acknowledgments

N.A. acknowledges support from NSF grant AST 1413319 as well
as NASA STScI grants GO 11686, 12022, 14242, 14054, and 14176, and NASA ADAP 48020.

Based on observations collected at the European Organisation for
Astronomical Research in the Southern Hemisphere
under ESO programmes 092.B-0267 (PI: Benn).


\bibliography{apj-jour,dsr-refs}

\end{document}